\documentclass[a4paper]{article}

\usepackage[left=1in,right=1in,top=1in,bottom=1in]{geometry}

\usepackage[numbers]{natbib}

\usepackage{amssymb}
\usepackage{amsthm}

\usepackage{url}
\usepackage{breakurl}
\usepackage{mathtools}
\usepackage{nicefrac}
\usepackage{xcolor}
\usepackage{siunitx}
\usepackage{todonotes}
\usepackage{subcaption}

\usepackage[vlined,linesnumbered]{algorithm2e}
\SetKw{Continue}{continue}
\SetKw{Break}{break}

\usepackage{mwe}
\newlength{\tempheight}
\newlength{\tempwidth}
\newcommand{\rowname}[1]%
{\rotatebox[origin=c]{90}{\makebox[\tempheight][c]{\textbf{#1}}}}
\newcommand{\columnname}[1]%
{\makebox[0.25\linewidth][c]{\textbf{#1}}}

\usepackage[switch]{lineno}

\begin{document}
\sloppy
\let\WriteBookmarks\relax
\def\floatpagepagefraction{1}
\def\textpagefraction{.001}

\title{Reconstructing the Dynamic Sea Surface from Tide Gauge Records Using Optimal Data-Dependent Triangulations}

\author{Alina Nitzke\footnote{nitzke@igg.uni-bonn.de} \and Benjamin Niedermann\footnote{niedermann@igg.uni-bonn.de} \and Luciana Fenoglio-Marc\footnote{fenoglio@geod.uni-bonn.de} \and J\"urgen Kusche\footnote{kusche@geod.uni-bonn.de} \and Jan-Henrik Haunert\footnote{haunert@igg.uni-bonn.de}}

\date{Institute of Geodesy and Geoinformation,\\ University of Bonn, Germany}

\maketitle

\begin{abstract}
Reconstructions of sea level prior to the satellite altimeter era are usually derived from tide gauge records; 
however 
most algorithms for this
assume that modes of sea level variability are stationary which is not true over several decades. Here we suggest a method that is based on optimized data-dependent triangulations of the network of gauge stations. Data-dependent triangulations are triangulations of point sets that rely not only on 2D point positions but also on additional data (e.g.\ elevation, anomalies). In particular, min-error criteria have been suggested to construct triangulations that approximate a given surface. In this article, we show how data-dependent triangulations with min-error criteria can be used to reconstruct 2D maps of the sea surface anomaly over a longer time period, assuming that height anomalies are continuously monitored at a sparse set of stations and, in addition, observations of a reference surface is provided over a shorter time period. At the heart of our method is the idea to learn a min-error triangulation based on the available reference data, and to use the learned triangulation subsequently to compute piece-wise linear surface models for epochs in which only observations from monitoring stations are given. Moreover, we combine our approach of min-error triangulation with $k$-order Delaunay triangulation to stabilize the triangles geometrically.
We show that this approach is in particular advantageous for the reconstruction of the sea surface by combining tide gauge measurements (which are sparse in space but cover a long period back in time) with data of modern satellite altimetry (which have a high spatial resolution but cover only the last decades). We show how to learn a min-error triangulation and a min-error $k$-order Delaunay triangulation using an exact algorithm based on integer linear programming. We confront our reconstructions against the Delaunay triangulation which had been proposed earlier
for sea-surface modeling and find superior quality. With real data for the North Sea we show that the min-error triangulation outperforms the Delaunay method significantly for reconstructions back in time up to 18 years, and the $k$-order Delaunay min-error triangulation even up to 21 years for $k=2$. With a running time of less than one second our approach would be applicable to areas with far greater extent than the North Sea.
\end{abstract}

\section{Introduction}
Observations of the 
spatially and temporally varying
sea surface height are important for 
understanding 
short-
and long-term  ocean processes like tides and currents, the response to atmospheric forcing, climate-driven variability or anthropogenic sea level rise (SLR). In addition, they are used to 
initialize short-term sea level forecasts and validate multi-decadal sea level predictions.

Satellite altimeters enable a global measurement of the sea surface over most of the ocean since the launch of the Topex/Poseidon satellite in 1992. 
Compared to tide gauges, satellite altimetry provides measurements of higher spatial resolution along the satellite track and of lower temporal resolution.
After a gridding procedure, altimetric data records such as the ESA Climate Change Initiative Sea Level (SLCCI) fields that we use here can be viewed as a mathematical representation of a reference (sea) surface via a discrete set of points. 

Another measurement of the sea surface is provided by land-based tide gauges, which can date back to the end of the 18th century. As the tide gauge stations are not uniformly distributed on the Earth's surface, observations are sparse in space. Before the satellite era, only tide gauge observations were available. To investigate past sea level and climate change, a long area-wide history of the sea surface is required. Aim of this paper is to reconstruct the dynamic sea surface at epochs where altimetric data is not available.

The conventional methodology for sea-level reconstruction is based on decomposing grids of altimetric (or model-based) sea level into 
spatial patterns and the corresponding temporal modes, and then fitting the dominating patterns in the pre-altimetry era to the tide gauge records. Patterns have been mostly derived through Empirical Orthogonal Function (EOF) analysis \cite{KaplanEtAl1998, ChurchEtAl2004, ChurchWhite2011},
or extensions such as rotated %
\citep{MeyssignacEtAl2012} or cyclostationary EOF \citep{StrassburgEtAl2014}.
However, the EOF method assumes that the covariance of the observations is constant, while the  physical modes of sea level variability are not stationary over many decades. They are also not orthogonal in a mathematical sense. Moreover, it is difficult to assign weights to individual tide gauge records, while their  heterogeneous spatial distribution tends to bias reconstructions.
This led recent studies to 
turn to probabilistic fingerprint techniques \cite{HayEtAl2015, DangendorfEtAl2017} or %
reconstruct sea level directly from multilinear regressions \citep{MadsenEtAl2019} or  Delaunay triangulations of tide gauges and %
interpolation \citep{OlivieriSpada2016}.

In fact, the latter paper proves the concept of triangulation-based sea level reconstructions by comparing to altimetric sea level, provided a suitable (GPS-colocated) tide gauge configuration encloses the considered ocean basin. Olivieri and Spada~\cite{OlivieriSpada2016} use altimetric data for assessing their reconstructions but work with a fixed triangulation which does not consider the sea surface signal. In contrast, here we suggest optimizing the configuration of the triangulation from the data in the overlap of the altimetry and tide gauge period, and then apply local interpolation in the pre-altimetry period.

In mathematical terms, we consider the locations of tide gauges as forming a point set on a two-dimensional manifold (such as the plane, the sphere, or an ellipsoid, depending on the region size), with the region of interest bounded by the convex hull of the point set. Sea surface height can be viewed as a function of time defined on the manifold, of which we have measured values for given epochs and gauge locations but possibly including gaps. 

\paragraph{Our contribution.} We present a new approach for reconstructing 
an (unobserved) area-wide sea surface from tide gauge records, given a limited span of overlap with observed area-wide altimetry grids. Our approach works via learning data-dependent triangulations based on tide gauges at epochs where area-wide sea-surface observations are given through satellite altimetry. These observations are used as reference surface for the computation of the triangulation. We consider both \emph{min-error triangulations} which minimize the misfit between the triangulation and the reference surface, as well as \emph{min-error higher-order Delaunay triangulations} which minimize the misfit between the triangulation and the reference surface subject to the constraint that the result must be a higher-order Delaunay triangulation~\cite{gudmundsson2002higher}. In the latter case we obtain triangulations that represent a compromise between pure min-error triangulations and  pure Delaunay triangulations, 
adopting the useful
properties of both concepts.   
Utilizing integer linear programming, 
an approach for solving generic combinatorial optimization problems, we construct optimal triangulations in the sense that %
no triangulations exist that satisfy the same constraints and that have a smaller error with respect to the reference surface.  

We transfer the learned triangulation to the same tide gauge stations at epochs
where no altimeter data is available, in order to reconstruct the dynamic sea surface %
by interpolating between the tide gauge measurements within each triangle. 
We show that the new approach achieves a significantly better dynamic sea-surface reconstruction using min-error (higher-order Delaunay) triangulations than by using Delaunay triangulations. For a skill metrics, we derive the variance reduction of the min-error (higher-order Delaunay) triangulation  compared to the Delaunay triangulation, both contrasted against the measured altimetric reference surface. 

The paper is organized as follows. First, we outline the formal model of our approach in Section~\ref{sec:formalModel}. Second in Section~\ref{sec:relatedWork} we give an overview of related work on triangulating point sets. In Section~\ref{sec:methodology}, we present a method for computing min-error (higher-order Delaunay) triangulations in more detail. Afterwards, in Section~\ref{sec:experiments} we provide an application to the reconstruction of sea level in the North Sea from tide gauges. Finally, we present a conclusion in Section~\ref{sec:conclusion}.

\section{Formal Model}\label{sec:formalModel}
For the representation of the sea surface through triangulations, we define the tide gauge locations as a set $P$ of $n$ points in the plane, $conv(P) \subseteq \mathbb{R}^2$ as its convex hull, and $h \colon P \rightarrow \mathbb{R}$ to be a function representing observed values for an epoch at each tide gauge $p \in P$.
Our aim is to define a piece-wise linear surface $s \colon conv(P) \rightarrow \mathbb{R}$ by triangulating $P$ and interpolating $h$ linearly within each triangle for the dynamic sea-surface reconstruction such that $h(p) = s(p) \quad \forall p \in P$. In our application, $s$ denotes a triangulation in 3D that defines sea surface anomalies (SSA). In general, $P$ can be any point set that is enriched with data values defined by any function $h$.

\begin{figure*}
\begin{subfigure}[c]{0.32\linewidth}
\includegraphics{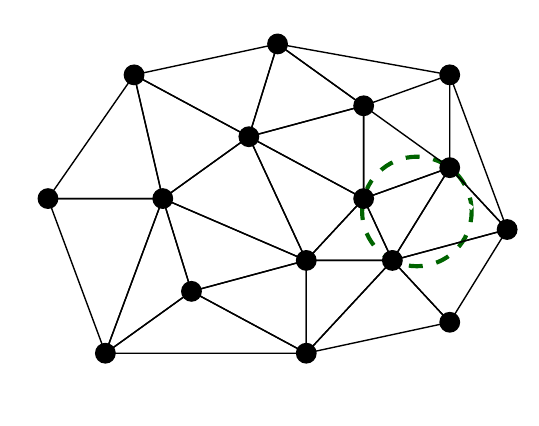}
\subcaption{}
\label{fig:triangulationModels_a}
\end{subfigure}
\begin{subfigure}[c]{0.32\linewidth}
\includegraphics{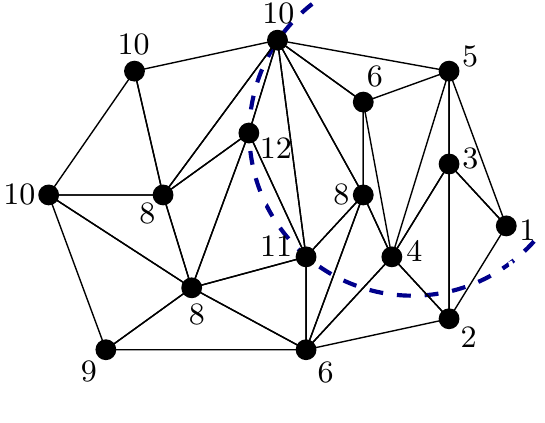}
\subcaption{}
\label{fig:triangulationModels_b}
\end{subfigure}
\begin{subfigure}[c]{0.32\linewidth}
\includegraphics{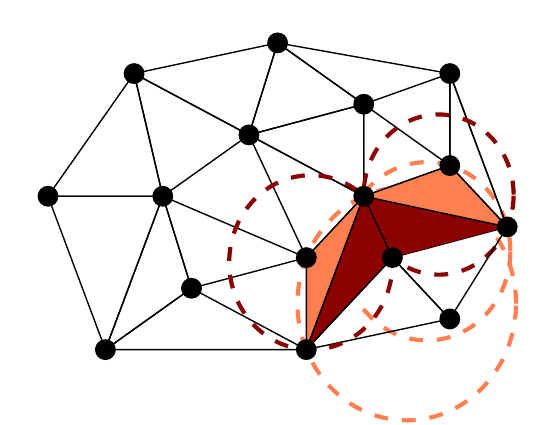}
\subcaption{}
\label{fig:triangulationModels_c}
\end{subfigure}
\caption{Comparison of three triangulation models based on the same point set. The dashed lines present the circumcircle of a triangle showing the circumcircle property of the corresponding triangulation method. (a) Delaunay triangulation that is defined by the empty-circle property. (b) Schematic illustration of a data-dependent triangulation that minimizes for each triangle the sum of variances of the data of its vertices. This triangulation does not consider a convenient triangle geometry so there can be an infinite number of points within the circumcircle of a triangle. (c) Higher-order Delaunay triangulation for $k=2$. Here, the red triangles are $1$-order Delaunay triangles while the orange triangles are $2$-order Delaunay triangles.}
\label{fig:triangulationModels}
\end{figure*}

Olivieri and Spada~\cite{OlivieriSpada2016} use Delaunay triangulations for representing the surface $s$. The Delaunay triangulation is defined by triangles whose circumcircle does not contain any other point of $P$, see Figure~\ref{fig:triangulationModels_a}. Its advantage is the convenient geometry of the triangles which are mostly \emph{well-shaped} meaning that the smallest angle of the triangles is maximized~\cite{Berg:2008:CGA:1370949}. However, the Delaunay triangulation entirely disregards data values.
Olivieri and Spada~\cite{OlivieriSpada2016} use altimeter data  to evaluate the quality of reconstruction through interpolation using $h$. 

In contrast to Olivieri~and~Spada~\cite{OlivieriSpada2016}, we suggest to include 
the data already in the construction of the
triangulation, i.e.\ prior to the actual interpolation and reconstruction step. For 
this, we apply \emph{data-dependent triangulations}, which were introduced by Dyn~et~al.~\cite{DynEtAl1990}. For an illustration see Figure~\ref{fig:triangulationModels_b}. The core idea of data-dependent triangulations is that they are computed using data values given for the point set $P$ which are defined by the function $h$, and under consideration of multiple criteria such that the given data is best represented by the resulting triangulation. However, we propose to use the altimeter data as reference data for the triangles in order to obtain the data-dependent triangulation in the optimization step. To this end, we minimize the sum of squared differences between reference values and interpolated observations within a triangle.

Accordingly, the problem is extended to the computation of the triangulation in such a way that $s$ \emph{approximates} a given function $f\colon \mathbb{R}^2 \to \mathbb{R}$  as good as possible, where in our case $f$ represents \emph{reference values} obtained from altimeter measurements of satellites. More precisely, we aim at minimizing $\iint_{\Omega} \xi ( s(x, y) - f(x, y) ) d{\Omega}$, where $\Omega \subseteq conv(P)$ is a given area of interest and $\xi \colon \mathbb{R} \to \mathbb{R}$ rates the 
misfit
between the interpolated value and the reference value at $(x,y)$. Assuming that $\Omega$ is represented by a discrete set $Q$ of points, the objective is to
\begin{linenomath*}
\begin{equation*}
\mathrm{minimize} \quad \sum_{(x,y) \in Q} \xi(s(x, y) - f(x, y)) \label{eq:obj}.
\end{equation*}
\end{linenomath*}
In our application we rate the error as $\xi(\delta) = \delta^2$, but we may also use any other 
metrics
depending on the specific application.

\begin{figure}
\center
\includegraphics{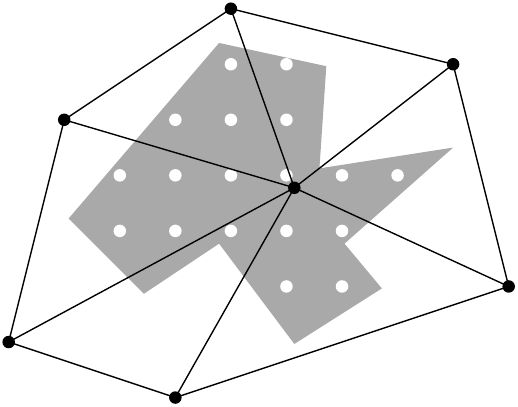}
\caption{A triangulation (black) approximating a surface defined by a set $Q$ of points (white dots) within a region of interest $\Omega$ (gray-shaded polygon).}
\label{fig:problem}
\end{figure}

Figure~\ref{fig:problem} illustrates this problem. We call the resulting triangulation \emph{min-error triangulation}. 

However, as the geometry of the triangles is not taken into account in this problem setting, this may result in \emph{badly-shaped} triangles, see Figure~\ref{fig:triangulationModels_b}, i.e.\ triangles that are long and narrow including an inner angle close to $180^\circ$, which can lead to high interpolation error. In order to obtain a triangulation which is defined by both well-shaped triangles and data-dependent criteria, we here suggest to combine data-dependent triangulations with \emph{higher-order Delaunay triangulations} \cite{gudmundsson2002higher}. The main idea is to allow at most $k$ points within the circumcircle of each triangle, see Figure~\ref{fig:triangulationModels_c}; we note that for a valid triangulation those $k$ points are not allowed in the triangle $\tau$ itself. In contrast to the Delaunay triangulation\footnote{A Delaunay triangulation is uniquely defined under the assumption that no more than three points lie on a common circle.} this triangulation is not uniquely defined, but with increasing $k$ the number of possible triangulations increases. The variable $k$ defines the \emph{order} of the triangulation and the corresponding triangulation is called \emph{$k$-order Delaunay triangulation}; we will call this briefly \emph{$k$-OD triangulation}. 
An example for a $2$-OD triangulation is shown in Figure~\ref{fig:triangulationModels_c}. Accordingly, the Delaunay triangulation can be interpreted as a $0$-OD triangulation, since no points are permitted in the circumcircle of a triangle. In contrast, a min-error triangulation is equivalent to an $\infty$-OD triangulation, i.e., the circumcircle of a triangle in the triangulation may contain the $n-3$ remaining points in $P$. Still we refrain from calling it a $\infty$-OD triangulation, but stick to the term \emph{min-error triangulation}.

As $k$-OD triangulations are not uniquely defined, we use this additionally obtained degree of freedom to optimize further data-dependent criteria. 
We call a $k$-OD triangulation $T$ a \emph{min-error $k$-order Delaunay triangulation} (short min-error \emph{$k$-OD} triangulation) if there is no other $k$-OD triangulation that approximates the reference surface better than $T$. As the number of possible candidate triangles for the triangulation increases with increasing order $k$, we can control the triangulation properties in terms of triangle geometry and the integration of data-dependent information by choosing $k$ appropriately: a small $k$ yields Delaunay-like triangles, while a large $k$ gives more freedom for optimization.

\section{Related Work on Triangulations} \label{sec:relatedWork}

Triangulating point sets is a fundamental problem of computational geometry \citep{Berg:2008:CGA:1370949}.
The Delaunay triangulation is often used for tasks concerning the modeling of a terrain surface, since it yields triangles that satisfy  multiple useful geometric criteria.
It can be characterized as seeking to maximize the minimum angle of all the angles of all the triangles in the triangulation \cite{Sibson1978}. Furthermore, it maximizes the mean inradius of the triangles \cite{l-dtmmi-CCCG94}.  Several efficient algorithms for computing the Delaunay triangulation are known \cite{Berg:2008:CGA:1370949,lee1980two,su1997comparison,Fortune2004}.
However, despite its high relevance, the Delaunay triangulation does not optimize all criteria that can be of interest.

Dyn~et~al.~\cite{DynEtAl1990} have introduced {data dependent triangulations} as triangulations that are computed under consideration of data values given for the point set $P$. The authors have considered multiple criteria, including {min-error criteria}, which are particularly relevant for our application. 

The Delaunay triangulation and the data-dependent triangulation can be considered as two 
end-members 
in the sense that either the focus 
is set on well-shaped triangles and the additional data is neglected, or high importance is attached to the information of additional data while one would accept a less well-shaped triangle geometry. To address this problem, Gudmundsson~et~al.~\cite{gudmundsson2002higher} have designed higher-order Delaunay triangulations. 
Rodr\'{\i}guez~and~Silveira~\cite{RodriguezSilveira2016} show in their first experiments on higher-order Delaunay triangulations, in the context of terrain approximation, that the use of this triangulation concept can lead to significant improvements over the Delaunay triangulation. 
Independently, de Kok~et~al.~\cite{DeKokEtAl2007} have used higher-order Delaunay triangulations in combination with two different heuristics in order to reduce the local minima in terrain models. Also Dyn et al.~\cite{DynEtAl1990} have tested an inexact 
approximate local-search method in the context of pure data-dependent triangulation, and other researchers have used similar heuristics~\citep{ALBOUL20001,WangEtAl2001}.

In contrast, our approach is based on exact optimization, to avoid heuristics and better exploit the potential of the method.
Little is known regarding the computational complexity of computing data dependent triangulations.
Van Kreveld et al.~\cite{vanKreveld2010} succeeded in developing an exact algorithm for first-order Delaunay triangulations optimizing a min-max criterion.

If no efficient algorithm for a problem is known, it is reasonable to approach the problem with integer linear programming (ILP).
ILP formulations solve an underlying problem by introducing integer variables, such that the objective function can be expressed as a linear combination of these variables constrained by linear inequalities \cite{Nemhauser1988}. 
Generally, an ILP can be expressed as follows:
\begin{linenomath*}
\begin{equation*}
\mathrm{Minimize} \quad \underline{c}^T \cdot  \underline{x}  \quad \mathrm{subject~to} \quad  \underline{A} \cdot  \underline{x} \le  \underline{b} \quad \mathrm{and} \quad  \underline{x} \in \mathbb{Z}^n \,,
\end{equation*}
\end{linenomath*}
where vectors $ \underline{c}$ and $\underline{b}$ as well as matrix $\underline{A}$ are given as input. 
Though the problem of solving an ILP is NP-hard \citep{GareyJohnson1990}, modern mathematical solvers 
can often solve problem instances of relevant size in reasonable time.
Kyoda~et~al.~\cite{KyodaEtAl1997} have presented an ILP-based method for the problem of finding a triangulation in the plane with the minimum total edge length.
More generally, de~Loera~et~al.~\cite{DeLoeraEtAl1996} have presented an integer programming formulation to model the 
set of all possible triangulations of a point set in a space of arbitrarily high dimension $d$.
Our method for computing a triangulation 
that best approximates a given surface is based on this approach.

\section{Algorithmic Approach} \label{sec:methodology}

\begin{figure*}
\centering
\includegraphics[scale=0.9]{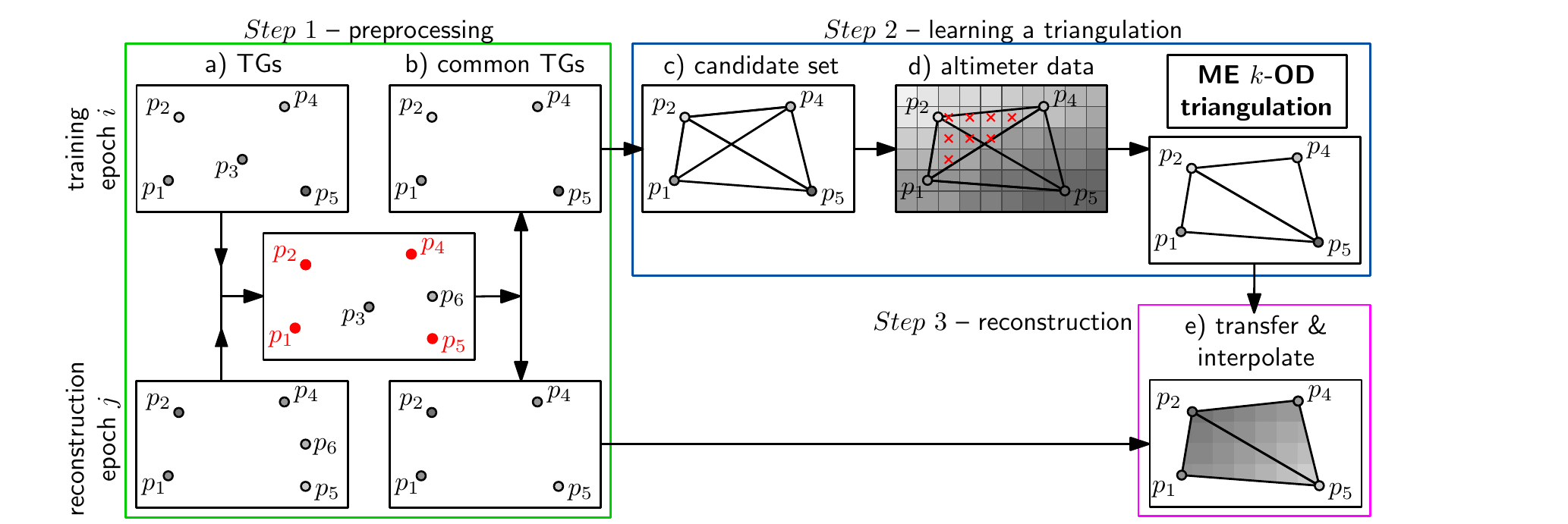}
\caption{Overview of our sea-surface reconstruction procedure at epoch $j$ by applying a min-error $k$-OD triangulation based on data of epoch $i$. The various shades of gray of the points show the different values of the observations for the tide gauge stations. The red points in the green box of \textit{Step 1} indicate the common points of both training and reconstruction epoch, which means that observations are available for these stations for both epochs. \emph{TGs} is short for tide gauges and \emph{ME $k$-OD triangulation} is short for min-error $k$-order Delaunay triangulation.}
\label{fig:method}
\end{figure*}

In this section we formalize the problem of dynamic sea-surface reconstruction. For the reconstruction of the dynamic sea surface at an epoch $j$ when only tide gauges exist, we use an optimized triangulation model learned on data of an epoch $i$ when altimeter data is available. We call epoch $i$ the \emph{training epoch} and epoch $j$ the \emph{reconstruction epoch}. 
Our approach consists of three steps; see also Figure~\ref{fig:method}.

\paragraph*{Step 1.} Some tide gauge stations do not provide measurements at all  epochs, rather the data may have gaps. Hence, we first determine all tide gauge stations that provide measurements for both training and reconstruction epoch, see the green box of Figure~\ref{fig:method}. We denote the resulting set by $P_{ij}$. This subsequently gives us the possibility of transferring the triangulation from the training epoch to the reconstruction epoch. For the running example we obtain $P_{ij}  =  \{p_1, p_2, p_4, p_5\}$; see Figure~\ref{fig:method} column a).

\paragraph*{Step 2.} For this step we compute the min-error $k$-OD triangulation $T_\mathrm{M}$ as follows, see the blue box in Figure~\ref{fig:method}. First, we create the set $\mathcal C$ of all possible triangles that might occur in a min-error $k$-OD triangulation; we call $\mathcal{C}$ \emph{candidate set} that consists of all \emph{candidate triangles}.
For the running example we obtain the candidate set
\begin{linenomath*}
\begin{equation*}
\mathcal{C} = \{\{p_1, p_2, p_4\}, \{p_1, p_2, p_5\}, \{p_1, p_4, p_5\}, \{p_2, p_4, p_5\} \}
\end{equation*}
\end{linenomath*}
as shown in Figure~\ref{fig:method} c).
Using integer linear programming, we compute the triangulation $T_\mathrm{M}$. To that end, we calculate for each triangle $\tau \in \mathcal{C}$ a cost that expresses the misfit of interpolated observations represented by $s$ with respect to the reference values given by $f$. As the altimetry measurements are given as gridded (i.e.\ raster) data, we assume that $f$ is defined for discrete coordinates $(x,y)$ each representing the center of a grid or raster cell; see the red crosses in triangle $\{p_1, p_2, p_4\}$ in Figure~\ref{fig:method} d). Our aim is to minimize the total cost of the triangles constituting $T_\mathrm{M}$. The calculation of the min-error $k$-OD  triangulation is explained in Section~\ref{sec:ILP} in greater detail.

\paragraph*{Step 3.} We transfer the min-error $k$-OD triangulation from the training epoch to the reconstruction epoch, in which we reconstruct the sea surface by interpolating $h$, see the magenta box in Figure~\ref{fig:method}.

\subsection{Integer Programming Formulation} \label{sec:ILP}

For our ILP formulation let $\mathcal{C}$ denote the set of candidate triangles.
We introduce a binary variable $x_\tau \in \{0,1\}$ for each $\tau \in \mathcal{C}$, which we interpret such that $x_\tau = 1$ if $\tau$ is selected for the resulting triangulation and $x_\tau=0$ otherwise.
It is easy to see that the size of $\mathcal{C}$ and, thus, the number of variables can be in the order of $n^3$.
However, the cubic size of $\mathcal{C}$ shall not be a major problem for point sets of modest size.

The objective function can be expressed with these variables as follows:
\begin{linenomath*}
\begin{align*} %
&\mathrm{Minimize} \quad \sum_{\tau \in \mathcal{C}} c(\tau) \cdot x_\tau\\
&\mathrm{where} \quad c(\tau) = \sum_{(x,y) \in Q \cap conv(\tau)}  \xi(s_\tau(x, y) - f(x, y))
\end{align*}
\end{linenomath*}
is the pre-computed cost of triangle $\tau \in \mathcal{C}$ and $s_\tau$ the plane induced by $\tau$.

To ensure that $\left\{\tau \in \mathcal{C} \mid x_\tau = 1\right\}$ is a valid triangulation, we introduce a set of linear constraints.
For this, let $E = \{ \{  u,v \} \mid u,v \in P, u \not= v \}$ be the set that contains an edge for each two distinct points.
Furthermore, for each edge $e \in E$, let $H_e^+$ and $H_e^-$ be the two half planes that are separated by the straight line extending $e$.
We denote with $E' \subseteq E$ the set of edges that form the convex hull of $P$ and 
assume $P \subseteq H_e^+$ for each edge $e \in E'$.

The first constraint of our ILP states that each edge that does not belong to the convex hull of $P$ must be on both sides incident to the same number of triangles:
\begin{linenomath*}
\begin{equation*}
\sum\limits_{\substack{\tau = \{p,q,r\} \in \mathcal{C} \colon \\ r \in P \cap H_e^+}} x_\tau = \sum\limits_{\substack{\tau = \{p,q,r\} \in \mathcal{C} \colon \\ r \in P \cap H_e^-}} x_{\tau}
\qquad \forall e = \{p ,q\} \in E \setminus E'
\end{equation*}
\end{linenomath*}
De~Loera~et~al.~\cite{DeLoeraEtAl1996}, while studying triangulations in arbitrary dimensions, refer to this as~\emph{interior cocircuit equation}
and show that with one additional equation, which can be of three different types, one can express the set of all triangulations of $P$.
In our approach, we use what de Loera~et~al.~\cite{DeLoeraEtAl1996} refer to as \emph{boundary cocircuit equations}.
That is, every edge of the convex hull is incident to exactly one selected triangle:
\begin{linenomath*}
\begin{equation*}
\sum\limits_{\substack{\tau = \{p,q,r\} \in \mathcal{C} \colon \\ r \in P \cap H_e^+}} x_\tau = 1 \qquad \forall e = \{p ,q\} \in E' . \label{eq:boundary-cocircuit}
\end{equation*}
\end{linenomath*}
According to the result by De~Loera~et~al.~\cite{DeLoeraEtAl1996}, it is sufficient to introduce this constraint %
for only one edge $e \in E'$. However, we generate one equation constraint for each edge of the convex hull to strengthen our formulation and improve the running time.

\section{Experiments} \label{sec:experiments}
In this section, we present the data used in our experiments and describe our evaluation method. Furthermore, we show the results of the comparison between the dynamic sea-surface reconstruction by our method and by the Delaunay triangulation. 
We apply our approach to data covering the North Sea with an area of about \SI{750000}{\square\km}.

\subsection{Data and Experimental Setup} \label{sec:data}
\begin{figure}
\centering
\begin{subfigure}{\linewidth}
\center
\includegraphics{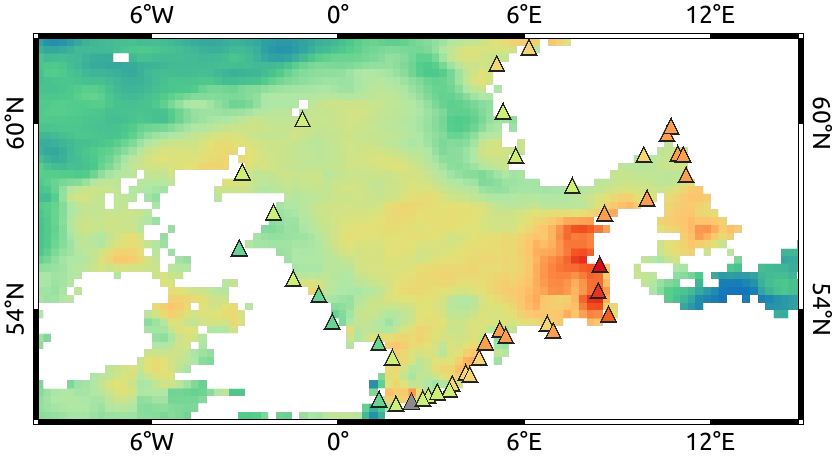}
\end{subfigure}
\begin{subfigure}{0.5\linewidth}
\includegraphics{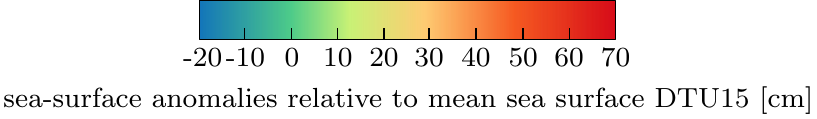}
\end{subfigure}
\caption{Area of interest for our experiments is the North Sea. The North Sea, a shallow shelf sea located between the UK, Central Europe and Scandinavia, opens to the Atlantic through the Norwegian Sea and through the English Channel and is connected to the Baltic via the Skagerrak. The data set consists of SSA obtained by tide gauges (triangles) and by satellite altimetry given by a grid (colored surface, here an example from January 1993). Gray filled triangles indicate that the respective tide gauge does not provide any observation for this epoch.}
\label{fig:data}
\end{figure}

Our approach combines tide gauge measurements with %
satellite altimetry. For the altimeter data we use the ESA Sea Level Climate Change Initiative (SLCCI) data base, which provides monthly gridded sea level anomalies derived from multi-satellite along-track observations \citep{ablain2015improved, esa}. The SLCCI grid resolution is 0.25 degree and the data record spans 22 years, from January 1993 to December 2015. All environmental and geophysical corrections were applied to the altimetric heights, including the dynamic  atmospheric correction (DAC) derived from the ERA Interim data set \cite{ablain2017satellite,legeais2018improved}.

The monthly tide-gauge time series are from the Permanent Service for Mean Sea Level (PSMSL) \cite{psmsl,holgate2012new}. 
We utilize %
only measurements which are available for more than 70\% of the time frame January 1993 to December 2015.
In the North Sea, the resulting data set consists of water levels for 276 months from 63 tide gauges and for the nearest altimeter data point. However, it should be mentioned that not all tide gauge provide uninterrupted time series, i.e.\ measurements for each epoch. This issue is considered in our method. 
For consistency with the altimeter data, the same DAC correction applied to the altimeter data is applied to the tide gauge records. While the altimeter time series are anomalies referenced to the mean sea surface (MSS) DTU15, each tide gauge record is relative to its own benchmark. Therefore, we de-mean each tide gauge record over 1993--2015 and add to the resulting anomalies the mean value of the nearest altimeter record, computed over the same time interval. In this way the transformed tide-gauge time series are anomalies relative to the mean sea surface DUT15 as well. As we are mainly interested in spatial patterns and as GNSS-based models tend to disagree with glacial isostatic modelling, we decided to not correct tide-gauge records for vertical land motion. We note this will lead to a systematic disagreement when we compare reconstructions to altimetry data.

Absolute sea level in the North Sea rose by 1.5 mm/yr over the 1900--2009 period \citep{WahlEtAl2013} and, depending on region, 1.3 to 3.9 mm/yr over the altimetry era (1993--2014) \citep{SterliniEtAl2017}. At time scales of decades, regional anomalies of sea level rates are dominated by 
sea level pressure and wind stress
but also local halo-%
and 
thermosteric effects play a role \citep{SterliniEtAl2017}. Rates are higher in the Eastern part 
as compared to the inner North Sea or the UK coast, which explains why mean altimetric sea levels do not match tide gauge means. 
An acceleration is visible in long tide gauge records, but studies have found that while %
recent rates %
were indeed high, they were not significantly higher than at other times in the 20th century (cf. \cite{WahlEtAl2013}). 

Seasonal sea level variability in the North Sea is mostly caused by the barotropic response to wind and pressure changes, and its decadal variability is closely related to the variability of atmospheric forcing \citep{PlagTsimplisl1999} and %
driven e.g.\ by variations of the North Atlantic Oscillation. A pronounced and spatially non-uniform low-frequency variability in the seasonal deviations from annual mean sea level exists \citep{FrederikseGerkema2018}; winter- %
and autumn-mean %
levels show highest values along the German~Bight.  %

The area of the North Sea is well sampled in the east, south and west by a total of 41 tide gauges marked by the triangles presented in Figure~\ref{fig:data}. The number of tide gauges that provide measurements for both training and reconstruction epochs varies between 29 and 41. Many studies found that time series of altimetric levels fit locally to gauge readings within 5--7 cm, better than high-resolution models (e.g.~\cite{SchallEtAl2016})

Tide gauge locations are provided in ellipsoidal coordinates; we map these to the plane using the Lambert Conformal Conic (LCC) projection (which is designed for the representation of relatively small areas in mid-latitudes). We use an existing implementation of the giCentre Utilities (\url{https://www.gicentre.net/software/utils/}).

\subsection{Evaluation Method of Reconstructions} \label{sec:quality}

\begin{figure*}
\centering
\includegraphics[scale=0.9]{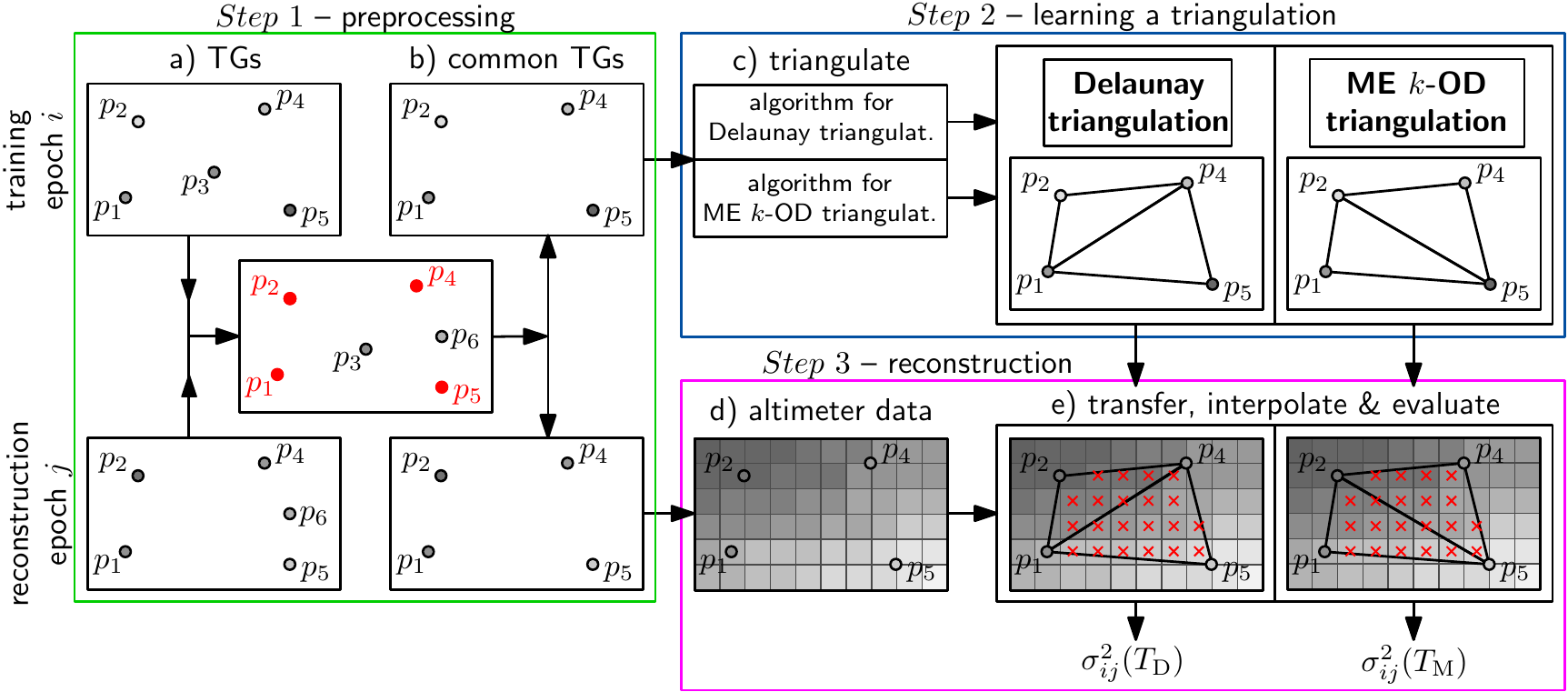}
\caption{For the evaluation we calculate the Delaunay triangulation $T_\mathrm{D}$ in addition to the min-error $k$-OD triangulation $T_\mathrm{M}$. We transfer both triangulations of the training epoch to the reconstruction epoch, interpolate with $h$ and evaluate by computing the variance between interpolated heights and reference heights given through altimeter data. Thus, we receive the variance for the Delaunay and the min-error $k$-OD triangulation. \emph{TGs} is short for tide gauges and \emph{ME $k$-OD triangulation} is short for min-error $k$-order Delaunay triangulation.}
\label{fig:flowchart}
\end{figure*}

For the evaluation, we compare the quality of the dynamic sea-surface reconstruction based on the min-error $k$-OD  triangulation~$T_\mathrm{M}$ to the quality of the reconstruction based on the Delaunay triangulation~$T_\mathrm{D}$. More specifically, we determine for $T_\mathrm{M}$ and $T_\mathrm{D}$ the misfit variances with respect to the altimeter data, respectively, and use them to assess the absolute variance reduction when applying $T_\mathrm{M}$ instead of $T_\mathrm{D}$. In order to systematically obtain the variance reduction for all given epochs, we apply four steps; the first three of them are illustrated in Figure~\ref{fig:flowchart}.

\paragraph*{Step 1.}
Analogous to the first step of our methodology (see Section~\ref{sec:methodology}) we determine the set $P_{ij}$ describing the tide gauge stations that provide measurements for both training and reconstruction epoch, see the green box in Figure~\ref{fig:flowchart}.

\paragraph*{Step 2.} We calculate a min-error $k$-OD triangulation $T_\mathrm{M}$ as well as a Delaunay triangulation $T_\mathrm{D}$ based on $P_{ij}$, see the blue box in Figure~\ref{fig:flowchart}.

\paragraph*{Step 3.} We transfer both $T_\mathrm{D}$ and $T_\mathrm{M}$ to the reconstruction epoch and reconstruct the sea surface through interpolating $h$ for both triangulations; see the magenta box in Figure~\ref{fig:flowchart}. To evaluate the reconstructions, we take the altimeter data at the reconstruction epoch serving as reference data. Afterwards, we compute the empirical variance of the reconstructions according to the formula for a general triangulation $T$:
\begin{linenomath*}
\begin{equation*} \label{eq:varianceij}
\sigma_{ij}^2(T) = \frac{1}{n-1}\sum_{\tau \in T} \sum_{(x,y) \in Q \cap conv(\tau)} \xi(s_{\tau}(x, y) - f_j(x, y)),
\end{equation*}
\end{linenomath*}
where $T$ is based on epoch $i$, $s_{\tau}$ is the plane defined by $\tau$ with respect to the tide gauges in epoch $j$, and $f_j(x,y)$ is the reference value at location $(x,y)$ for the reconstruction epoch~$j$.
Thus, with this we obtain the variance~$\sigma_{ij}^2(T_\mathrm{M})$ for the min-error $k$-OD  triangulation $T_\mathrm{M}$ and  the variance $\sigma_{ij}^2(T_\mathrm{D})$ for the Delaunay triangulation $T_\mathrm{D}$.

\paragraph*{Step 4.}
Finally, we compute for the min-error $k$-OD triangulation~$T_\mathrm{M}$
its \emph{variance reduction} with respect to the Delaunay triangulation~$T_\mathrm{D}$: 
\begin{equation*}
\Delta \sigma_{ij}^2 = \sigma_{ij}^2(T_\mathrm{M}) - \sigma_{ij}^2(T_\mathrm{D}). 
\end{equation*}
For $\Delta \sigma_{ij}^2 < 0$, our method achieves a better sea-surface reconstruction than the Delaunay triangulation. In contrast, if $\Delta \sigma_{ij}^2 > 0$, the Delaunay triangulation is a better reconstruction.

These four evaluation steps are applied on all possible combinations of training and reconstruction epochs within the provided time series from January 1993 to December 2015. More precisely, for a fixed training epoch $i$ the triangulation $T_\mathrm{M}$  is evaluated on all epochs computing $D$ variance reductions. Thus, we obtain $D^2$ variance reductions in total.

The variance reductions metrics allow us to assess the quality of a single reconstruction based on a min-error $k$-OD triangulation for a given training epoch $i$ and a reconstruction epoch $j$. However, they cannot directly be used to systematically assess how far back in the past a reconstruction epoch may lie on average such that the min-error $k$-OD  triangulation is the better choice than the Delaunay triangulation. In other words, how far back in time would we really benefit from using triangulations that were optimized at least partly to aid fitting best to the more recent reference data? In the following we consider that a reconstruction for a training epoch $i$ and a reconstruction epoch $j$ \emph{spans} $|i-j|$ epochs. We define an additional measure that rates the average quality of all reconstructions that span the same number of epochs. More specifically, for a given number $\Delta d\in \mathbb N$ let $\mathcal D(\Delta d)$ be the set of the variance reductions $\Delta \sigma^2_{ij}$ of all reconstructions that span $\Delta d$ epochs, i.e.,  $|i-j|=\Delta d$. In our evaluation we then consider the average variance reduction $q(\Delta d)$ for the reconstructions that span $\Delta d $ epochs, i.e.,
\begin{linenomath*} 
\begin{equation} \label{eq:quality}
q(\Delta d) = \frac{1}{|\mathcal D(\Delta d)|} \sum_{\Delta \sigma^2 \in \mathcal D(\Delta d)}  \Delta \sigma^2 .
\end{equation}
\end{linenomath*}
For $q(\Delta d) < 0$, our method achieves a better surface reconstruction than the Delaunay triangulation. Otherwise, for $q(\Delta d) > 0$  the Delaunay triangulation results in a better reconstruction.

In our use-case, one would only use reconstruction epochs that lie in the past, but in order to increase the amount of test cases used in our evaluation we also consider reconstruction epochs that occur after the training epochs.

\subsection{Results} \label{sec:results}

\begin{figure*}
    \centering
    \setlength{\tempwidth}{.4\linewidth}
    \settoheight{\tempheight}{\includegraphics[]{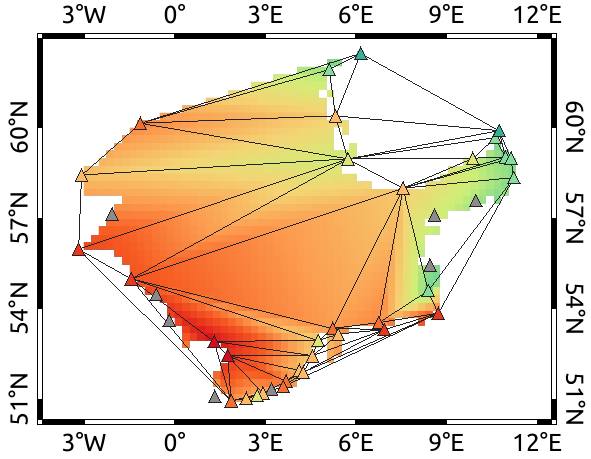}}%
    \hspace{\baselineskip}
    \columnname{min-error $2$-OD triangulation}\hfil
    \columnname{Delaunay triangulation}\\
    \rowname{June 2014}
    \begin{subfigure}{\tempwidth}
        \center
        \includegraphics{figs/results/reconstructions/201406_ILP.pdf}
        \vspace{-0.5\baselineskip}
        \caption[]{} 
        \label{fig:reconstruction_a}
    \end{subfigure}
    \begin{subfigure}{\tempwidth}
        \center
        \includegraphics{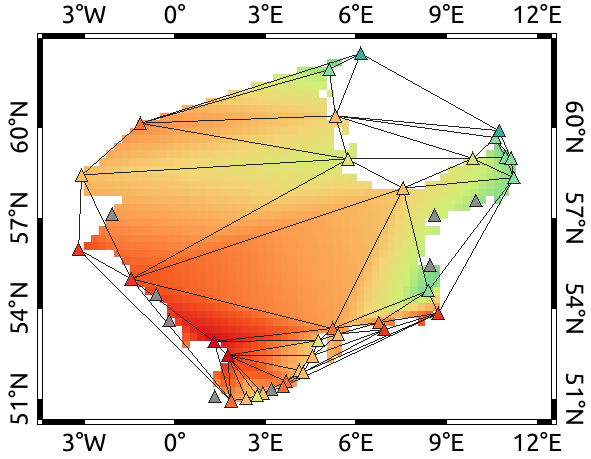}
        \vspace{-0.5\baselineskip}
        \caption[]{} 
        \label{fig:reconstruction_b}
    \end{subfigure}\\
    \rowname{June 2005}
    \begin{subfigure}{\tempwidth}
        \center
        \includegraphics{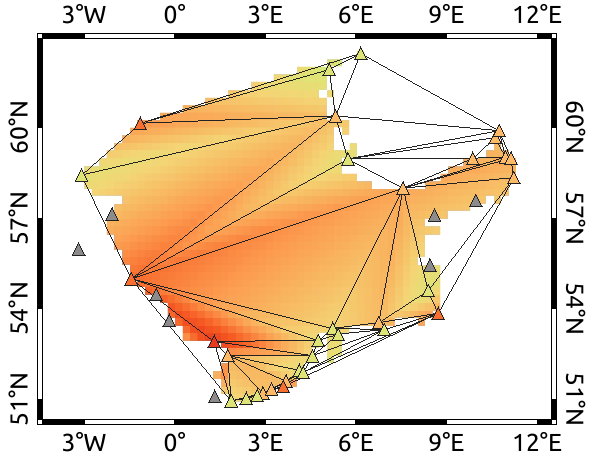}
        \vspace{-0.5\baselineskip}
        \caption[]{} 
        \label{fig:reconstruction_e}
    \end{subfigure}
    \begin{subfigure}{\tempwidth}
        \center
        \includegraphics{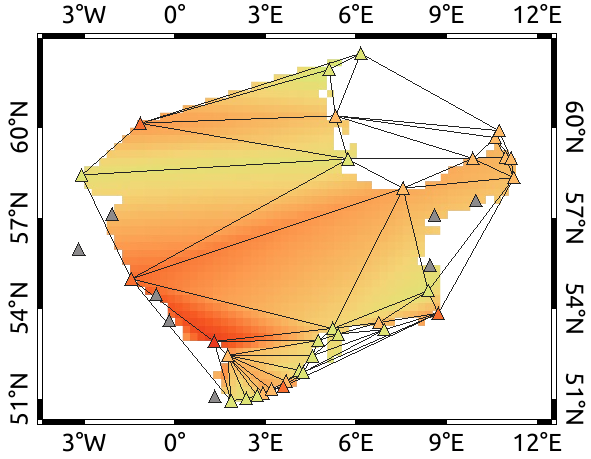}
        \vspace{-0.5\baselineskip}
        \caption[]{} 
        \label{fig:reconstruction_f}
    \end{subfigure} \\
    \rowname{June 2000}
    \begin{subfigure}{\tempwidth}
        \center
        \includegraphics{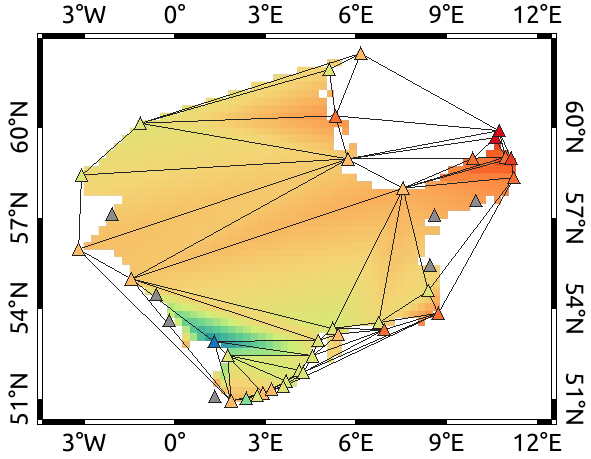}
        \vspace{-0.5\baselineskip}
        \caption[]{} 
        \label{fig:reconstruction_g}
    \end{subfigure}
    \begin{subfigure}{\tempwidth}
        \center
        \includegraphics{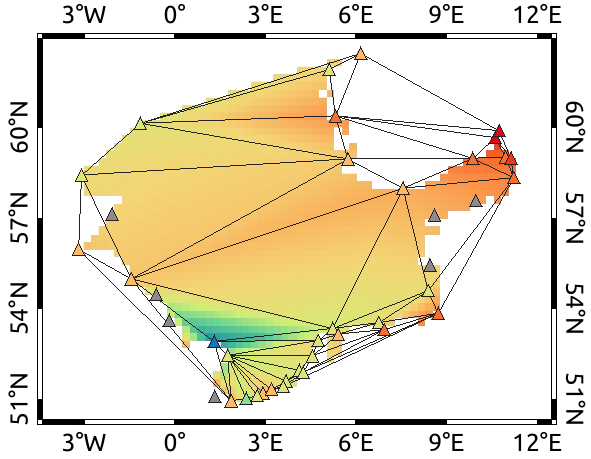}
        \vspace{-0.5\baselineskip}
        \caption[]{} 
        \label{fig:reconstruction_h}
    \end{subfigure} \\
    \rowname{June 1995}
    \begin{subfigure}{\tempwidth}
        \center
        \includegraphics{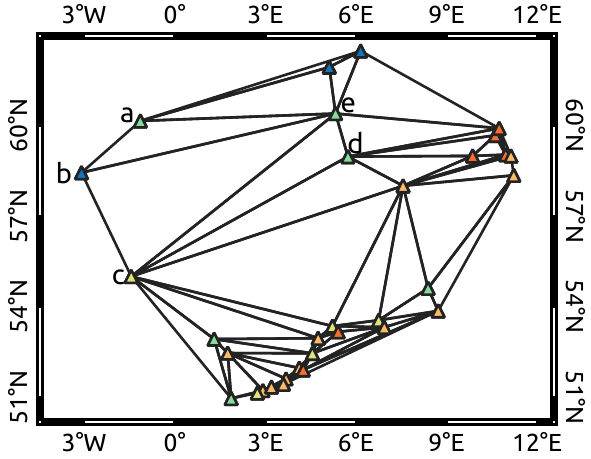}
        \vspace{-0.5\baselineskip}
        \caption[]{} 
        \label{fig:reconstruction_c}
    \end{subfigure}
    \begin{subfigure}{\tempwidth}
        \center
        \includegraphics{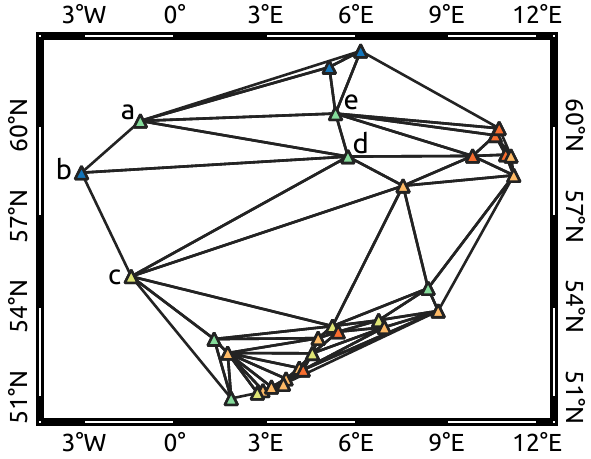}
        \vspace{-0.5\baselineskip}
        \caption[]{} 
        \label{fig:reconstruction_d}
    \end{subfigure}\\
    \begin{subfigure}{\linewidth}
        \center
        \includegraphics{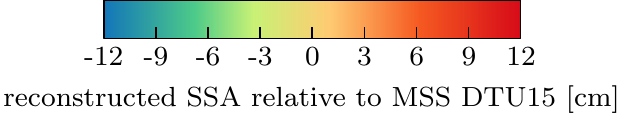}
    \end{subfigure}
    \caption{Learned triangulations based on the training epoch $i =$ June 2015 and the resulting reconstructions in cm relative to mean sea surface DTU15 for various reconstruction epochs. Left column: results for the min-error $2$-OD triangulation. Right column: results using the Delaunay triangulation. The reconstruction and the observation values of the tide gauge stations (triangles) are colored according to the given color bar (unit cm). Tide gauge stations that are not available for the training and reconstruction epoch are marked by gray triangles. }
    \label{fig:reconstruction}
\end{figure*}

\begin{figure*}
    \centering
    \setlength{\tempwidth}{.4\linewidth}
    \settoheight{\tempheight}{\includegraphics[]{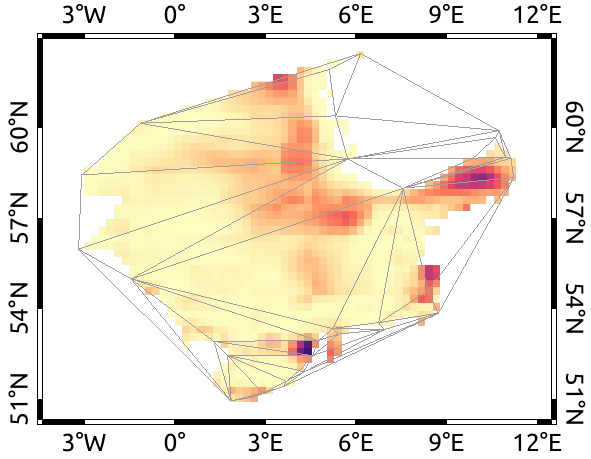}}%
    \centering
    \hspace{\baselineskip}
    \columnname{min-error $2$-OD triangulation}\hfil
    \columnname{Delaunay triangulation}\\
    \rowname{June 2014}
    \begin{subfigure}[c]{\tempwidth}
        \center
        \includegraphics[scale=1]{figs/results/anomaliesSquared/201406_ILP.pdf}
        \vspace{-0.5\baselineskip}
        \caption[]{} 
        \label{fig:misfitsSq_a}
    \end{subfigure}
    \begin{subfigure}[c]{\tempwidth}
        \center
        \includegraphics[scale=1]{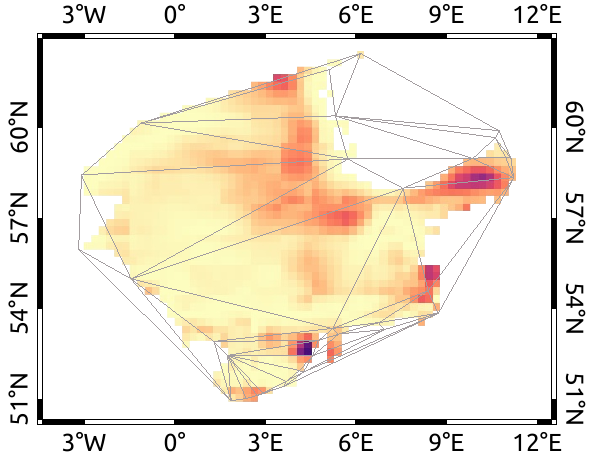}
        \vspace{-0.5\baselineskip}
        \caption[]{} 
        \label{fig:misfitsSq_b}
    \end{subfigure}\\
    \rowname{June 2005}
    \begin{subfigure}[c]{\tempwidth}
        \center
        \includegraphics[scale=1]{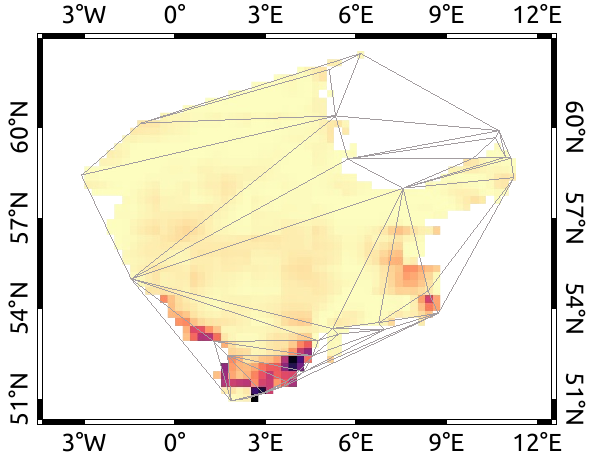}
        \vspace{-0.5\baselineskip}
        \caption[]{} 
        \label{fig:misfitsSq_e}
    \end{subfigure}
    \begin{subfigure}[c]{\tempwidth}
        \center
        \includegraphics[scale=1]{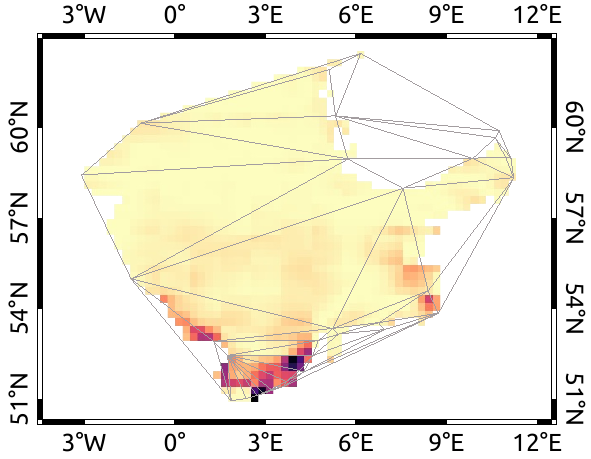}
        \vspace{-0.5\baselineskip}
        \caption[]{} 
        \label{fig:misfitsSq_f}
    \end{subfigure} \\
    \rowname{June 2000}
    \begin{subfigure}[c]{\tempwidth}
        \center
        \includegraphics[scale=1]{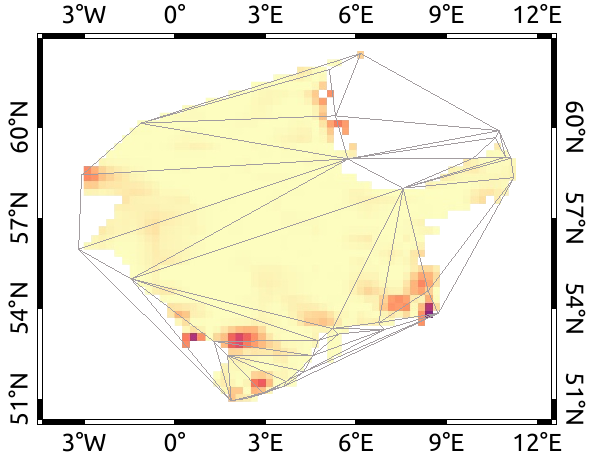}
        \vspace{-0.5\baselineskip}
        \caption[]{} 
        \label{fig:misfitsSq_g}
    \end{subfigure}
    \begin{subfigure}[c]{\tempwidth}
        \center
        \includegraphics[scale=1]{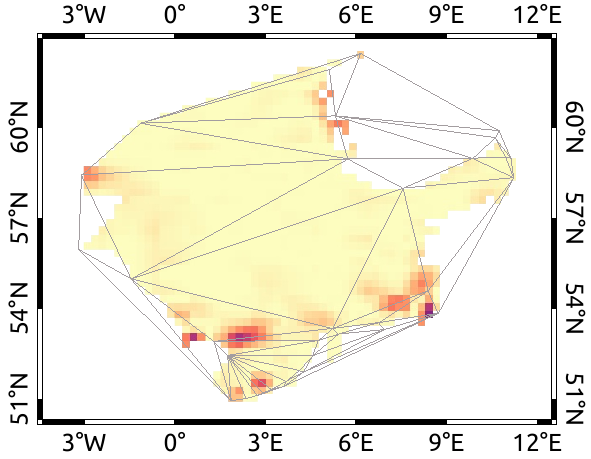}
        \vspace{-0.5\baselineskip}
        \caption[]{} 
        \label{fig:misfitsSq_h}
    \end{subfigure} \\
    \rowname{June 1995}
    \begin{subfigure}[c]{\tempwidth}
        \center
        \includegraphics[scale=1]{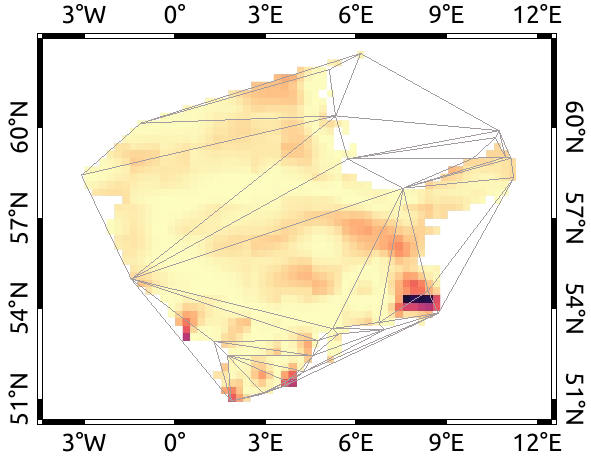}
        \vspace{-0.5\baselineskip}
        \caption[]{} 
        \label{fig:misfitsSq_c}
    \end{subfigure}
    \begin{subfigure}[c]{\tempwidth}
        \center
        \includegraphics[scale=1]{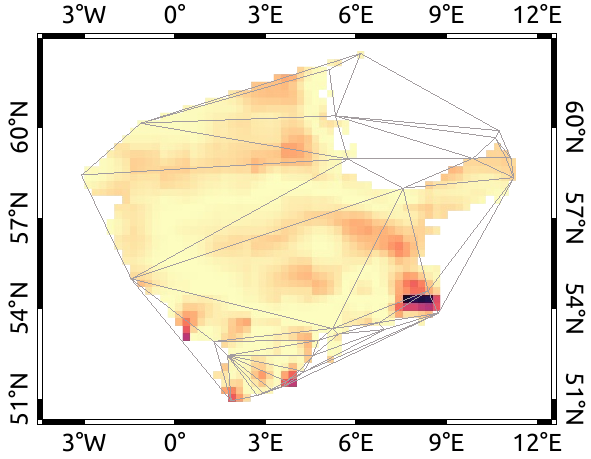}
        \vspace{-0.5\baselineskip}
        \caption[]{} 
        \label{fig:misfitsSq_d}
    \end{subfigure}\\
    \begin{subfigure}[c]{\linewidth}
        \center
        \includegraphics[scale=1]{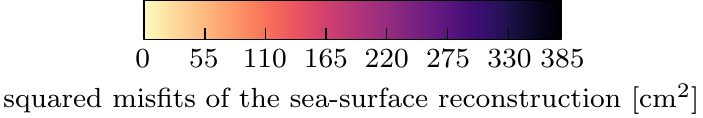}
    \end{subfigure}
    \caption[]{Squared misfits of the sea-surface reconstructions shown in Figure \ref{fig:reconstruction} and the reference altimeter data of the respective reconstruction epoch in $\text{cm}^2$. The training epoch is June 2015. Left column: squared misfits using the min-error $2$-OD triangulation for the reconstruction. Right column: squared misfits using the Delaunay triangulation for the reconstruction.}
    \label{fig:misfitsSq}
\end{figure*}

\begin{figure*}
    \centering
    \setlength{\tempwidth}{.4\linewidth}
    \settoheight{\tempheight}{\includegraphics[]{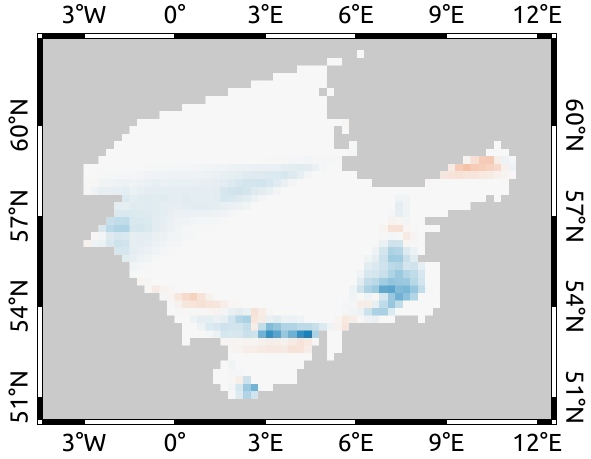}}%
    \centering
    \begin{subfigure}[c]{\tempwidth}
        \center
        \includegraphics[scale=1]{figs/results/diffsASq/201406.pdf}
        \vspace{-0.5\baselineskip}
        \caption[]{June 2014} 
        \label{fig:diffMSq_a}
    \end{subfigure}
    \begin{subfigure}[c]{\tempwidth}
        \center
        \includegraphics[scale=1]{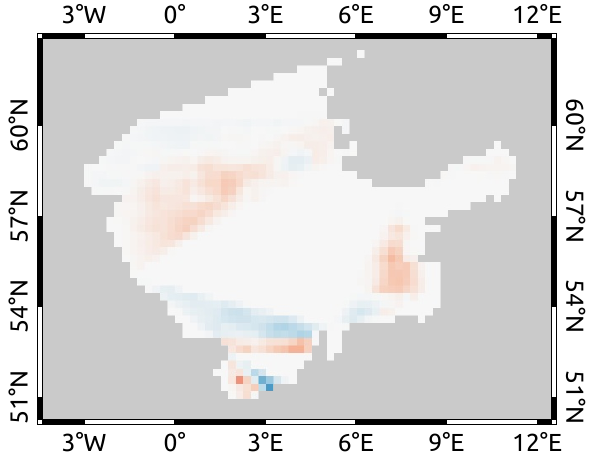}
        \vspace{-0.5\baselineskip}
        \caption[]{June 2005} 
        \label{fig:diffMSq_c}
    \end{subfigure}
    \begin{subfigure}[c]{\tempwidth}
        \center
        \includegraphics[scale=1]{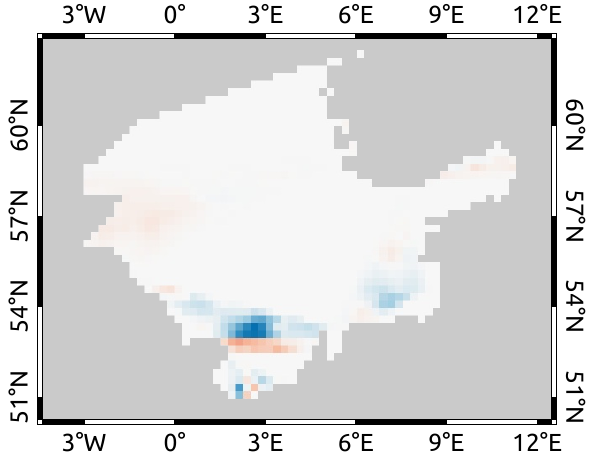}
        \vspace{-0.5\baselineskip}
        \caption[]{June 2000} 
        \label{fig:diffMSq_d}
    \end{subfigure}
    \begin{subfigure}[c]{\tempwidth}
        \center
        \includegraphics[scale=1]{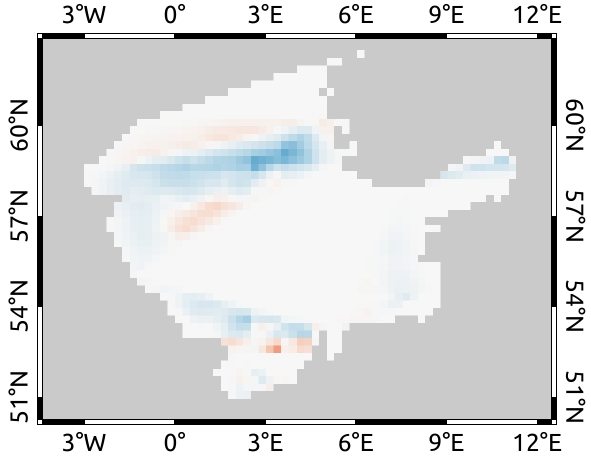}
        \vspace{-0.5\baselineskip}
        \caption[]{June 1995} 
        \label{fig:diffMSq_b}
    \end{subfigure}
    \begin{subfigure}[c]{\linewidth}
        \center
        \includegraphics[scale=1]{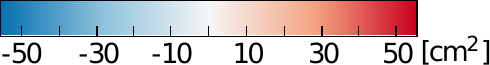}
    \end{subfigure}
    \caption[]{Difference of the squared misfits obtained by the min-error $2$-OD and the Delaunay triangulation (see Figure ~\ref{fig:misfitsSq}) of the respective reconstruction epoch in cm$^2$. The training epoch is June 2015. Regions colored blue (negative values) signify that our method reaches a better reconstruction than using the Delaunay triangulation. In contrast, regions that are colored red (positive values) indicate a better reconstruction through the Delaunay triangulation.}
    \label{fig:diffMSq}
\end{figure*}

\begin{figure*}
\center
\includegraphics{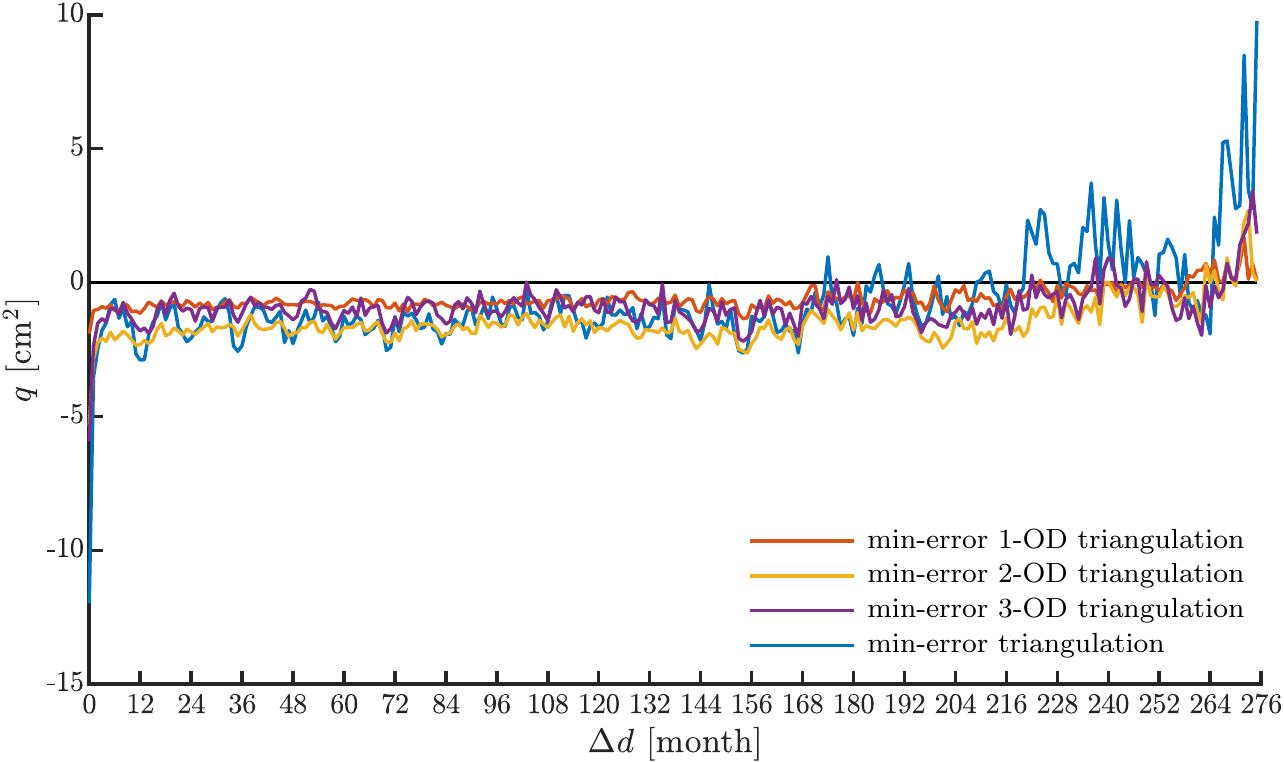}
\caption{Averaged variance reductions $q(\Delta d)$ of min-error $k$-OD triangulations compared to Delaunay triangulation computed from Eq.~\ref{eq:quality} for different $\Delta d$. The min-error $k$-OD  triangulations achieves better surface reconstructions than the Delaunay triangulation if $q(\Delta d) < 0$, otherwise  $q(\Delta d) > 0$.}
\label{fig:results}
\end{figure*}

\begin{figure*}
\center
\includegraphics{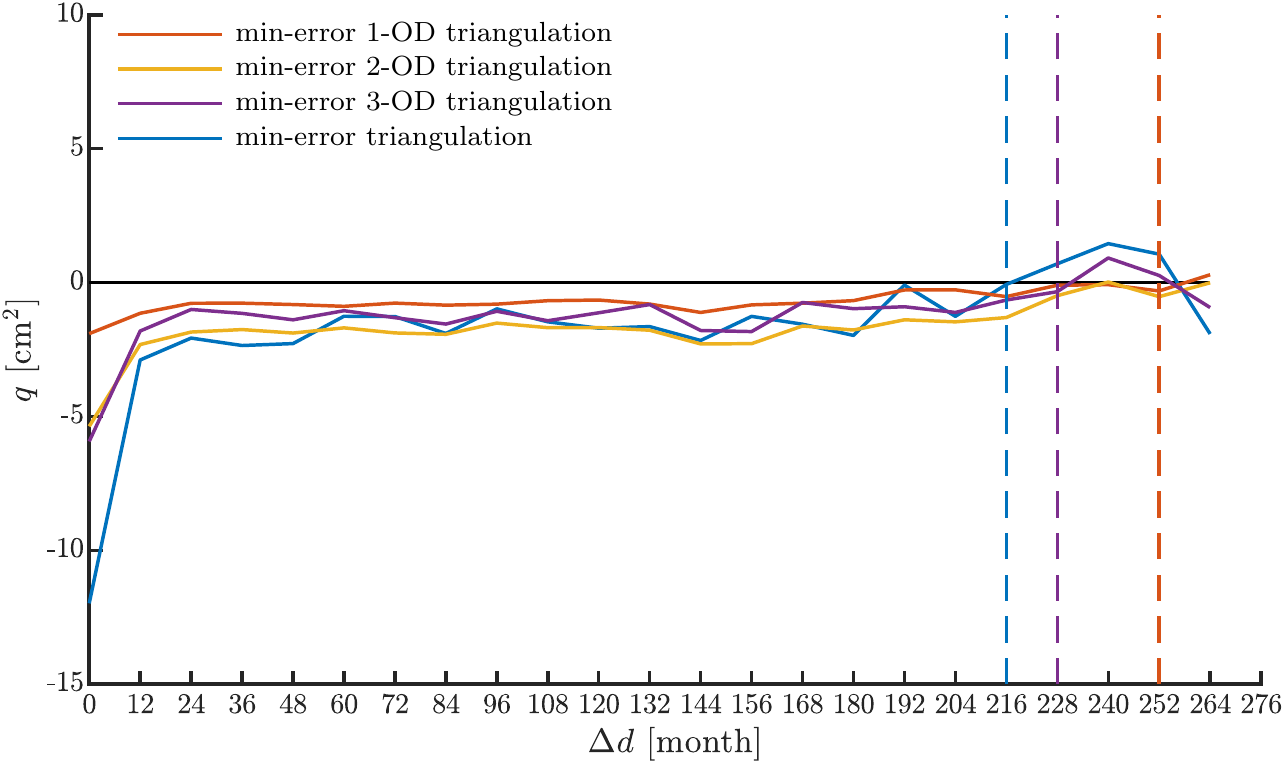}
\caption{Averaged variance reductions $q(\Delta d)$ of min-error $k$-OD  triangulations compared to Delaunay triangulations computed from Eq.~\ref{eq:quality} for climatological triangulations. The min-error $k$-OD triangulations achieves better surface reconstructions than the Delaunay triangulation if $q(\Delta d) < 0$, otherwise $q(\Delta d) > 0$. The dashed vertical lines indicate the largest $\Delta d$ for which all reconstructions with this $\Delta d$ or smaller are better than the Delaunay triangulation.}
\label{fig:results_yearly}
\end{figure*}

\begin{figure*}
\center
\includegraphics{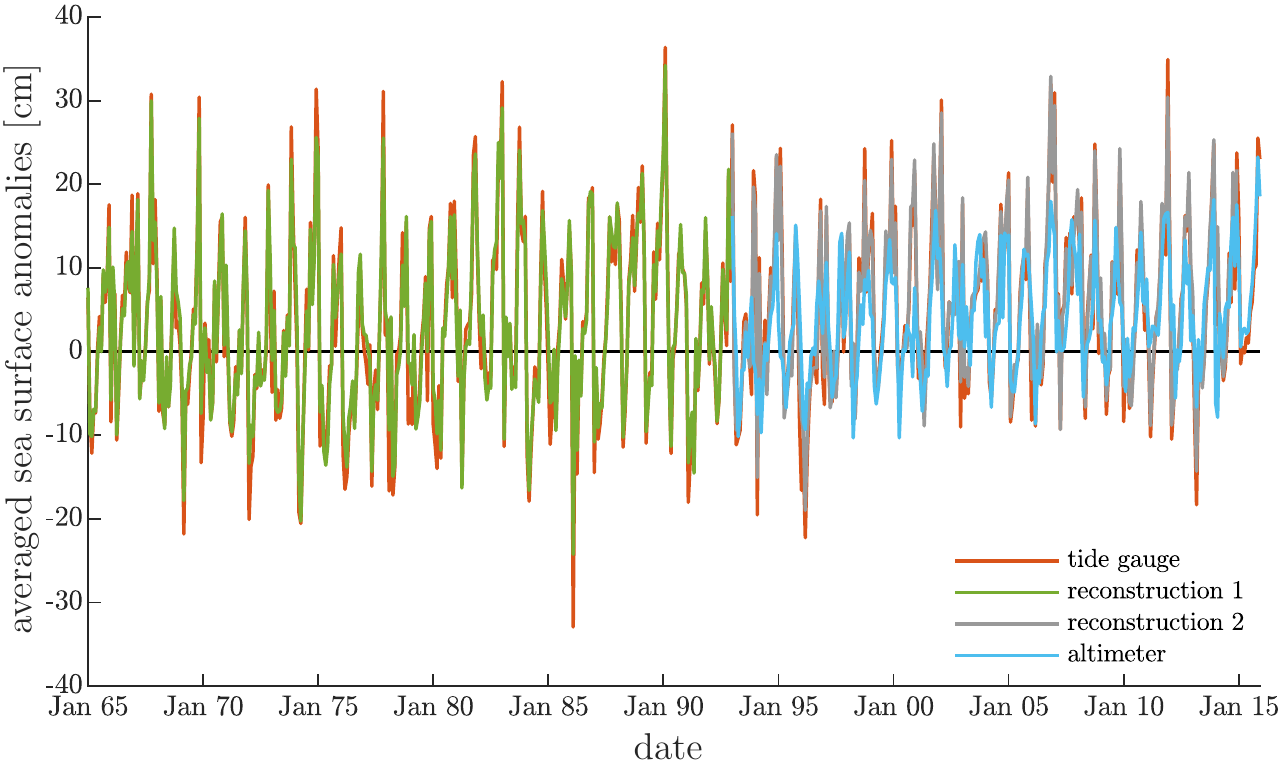}
\caption{Averaged SSA of the tide gauges (without DAC), altimeter data (with DAC) and the reconstructions based on min-error $2$-OD triangulations for the time period January 1965 to December 2015. Here, \emph{reconstruction 1} (green line) defines the reconstruction of the SSA based on trained triangulations with data of 1993 back to 1965, whereas \emph{reconstruction 2} (gray line) is the reconstruction of the SSA based on trained triangulations with data of 2015 back to 1993.}
\label{fig:means_LatWeighted}
\end{figure*}

\begin{figure*}
\center
\includegraphics{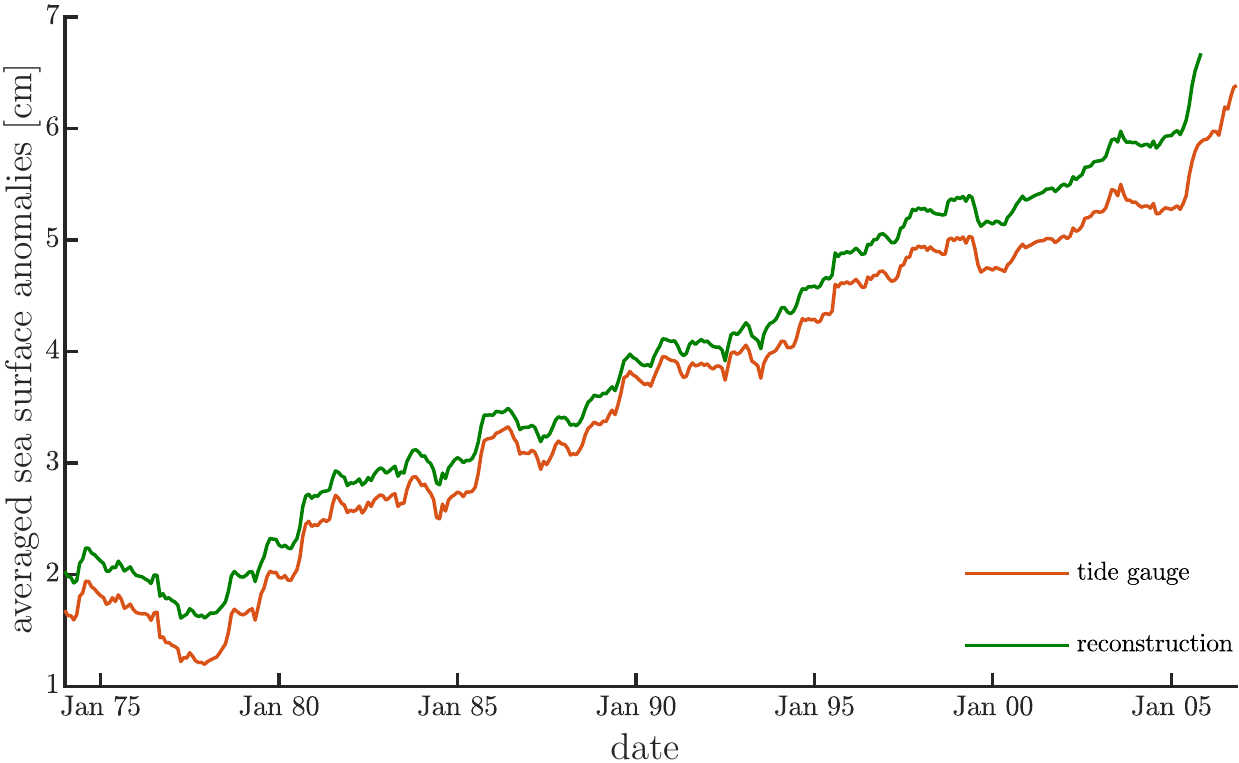}
\caption{Averaged SSA of the tide gauges, altimeter data and the reconstructions based on min-error $2$-OD triangulation filtered by a 19-years moving average for the time period January 1974 to December 2006. Here, the result for the \emph{reconstruction} (green line) is computed from the combination of the two reconstructions 1 and 2 of Figure~\ref{fig:means_LatWeighted}.}
\label{fig:movingAverage}
\end{figure*}

Figure~\ref{fig:reconstruction} shows exemplary results for triangulations that were learned from the training epoch June 2015 and applied on the reconstruction epochs June 1995, June 2000, June 2005, and June 2014. 
More precisely, for each reconstruction epoch we compare the results of the min-error $2$-OD triangulation, left column, and of the Delaunay triangulation, right column. We have chosen the min-error $2$-OD triangulation, because  it outperforms the other min-error $k$-OD triangulations as we will see later in the evaluation. 
We observe that the min-error $2$-OD triangulation and the Delaunay triangulation differ in the generated triangles: the min-error $2$-OD triangulation consists of significantly narrower triangles in some areas, while the Delaunay triangulation uses compact triangles. Consequently, comparing the reconstruction of the min-error $2$-OD with the reconstruction of the Delaunay triangulation shows obvious differences; for example consider the sub-triangulation of the tide gauges $a$, $b$, $c$, $d$ and $e$ in reconstruction epoch June 1995.

Moreover, Figure~\ref{fig:misfitsSq} shows the squared misfits between the reconstructions and the SSA reference of the respective reconstruction epoch. We  observe that both the min-error $2$-OD triangulation and the Delaunay triangulation yield results that are close to $0$ for most of the sea surface (yellow area). Further, 
visually it appears as if
they yield almost identical misfits. However, the differences of the squared misfits between both triangulations show that the  SSA is approximated better by the min-error $2$-OD triangulation than by the Delaunay triangulation for June 2014, June 2000 and June 1995; see Figure~\ref{fig:diffMSq}.  For June 2005 the Delaunay triangulation is the better choice; we assume that the training epoch does not suit the reconstruction epoch well. 

Figure~\ref{fig:results} presents the resulting averaged variance reductions $q(\Delta d)$ for each $\Delta d$ in the range between $0$ and $275$ months. 
First of all, the min-error triangulation exhibits periodic fluctuations with local minima approximately every $12$ months. As expected we observe a strong annual signal in the reconstruction. The curves of the averaged variance reduction for min-error $k$-OD triangulations with $k=1,2,3$ are significantly smoother than that of the min-error triangulation. The local minima that reoccur after each year are no longer that distinctive. This indicates that the triangulation is no longer only oriented to the altimeter data, but that the geometry of the triangles, i.e.\ their compactness, is also taken into account by the type of candidate triangles. Altogether, the results provide us with a simple approach to choose the training epoch $i$ for learning a min-error triangulation for reconstruction epoch $j$: reconstruct the dynamic sea surface for epoch $j$ using a min-error $k$-OD triangulation of epoch $i$ such that $i = j \pm 12 K $. We call this a \emph{climatological triangulation}.
Accordingly, we restrict the further analysis to the average variance reduction for $\Delta d = 12 K$ with $K\in \mathbb N$; see Figure~\ref{fig:results_yearly}. 

Next, we compare the min-error triangulation with the Delaunay triangulation in detail; see Figure~\ref{fig:results_yearly}. Using a min-error triangulation we achieve a better reconstruction than with a Delaunay triangulation for about $\Delta d \leq 216$ months.  Using the min-error $k$-OD  triangulation with $k \leq 3$ we can reconstruct the SSA better for an even longer period of time $\Delta d$, i.e.\ $228$ months ($=19$ years) for $k=3$, and $252$ ($=21$ years) for $k=2$ and $k=1$. Moreover, the min-error $2$-OD  triangulation outperforms the min-error $1$-OD  triangulation as it achieves a higher averaged variance reduction over $\Delta d$.

For %
longer reconstructions ($\Delta d > 228$ for $k=3$ and $\Delta d > 252$ for $k=1,2$), we cannot 
be certain 
which method produces better results. The positive variance reductions may also result from the fact that for large $\Delta d$ the variance reduction only takes few variance differences into account due to the lack of data. 
Further, we also have computed min-error $k$-OD  triangulations for $k > 3$, but initial experiments showed that the min-error $2$-OD  triangulation clearly prevailed min-error $k$-OD  triangulations for $k>3$. 

Finally, we reconstructed five decades of gridded SSA; thus including a significant period prior to the altimetry era. To this end, we used monthly tide gauge and altimeter data from 1993 onwards in combination with a climatological triangulation to reconstruct the SSA for years before 1993 back to 1965.
Figure~\ref{fig:means_LatWeighted} shows the mean SSA computed from tide gauge data that are available for the entire period January 1965 to December 2015 (without DAC), for the SSA reference given by altimeter data (provided from January 1993 to December 2015 with DAC) and  for the new reconstruction from January 1965 to December 1992.
We first observe that the variability of the altimeter area-mean appears similar when compared to the tide gauge mean in the January 1993 to December 2015 time frame, although peak levels are generally less pronounced (which is expected since near-shore variability is generally larger and the altimeters do not exactly cover the gauge locations). Note that even strong storms such as Xaver in 12/2013 average out to a large extend in monthly records. We notice a very similar  correlation also between the reconstruction and the tide gauges in the time range from January 1965 to December 1992. This supports our hypothesis that the presented approach can be used to reliably reconstruct the SSA based on time gauges for long-term periods. From our new reconstruction approach, we find a rate of area-mean sea level rise of 1.3 mm/yr over 1965--2015 and 2.2 mm/yr over {1993--2015} (altimetry 2.5 mm/yr).  The arithmetic mean rate from the gauges, over 1965--2015, points to a rate of 1.2~mm/yr, also very close to the area mean, but we suspect this may be due to chance here and may change with tide gauge selection. The small difference can be explained by the spatial variation of SLR in the North Sea. These results underpin the importance of having robust and reliable algorithms for the reconstruction of SSA maps from gauge data. Furthermore, we derive 19-year averages, in order to focus on short- and long-term interannual variability; see Figure~\ref{fig:movingAverage}. Periods of low and high water levels are broadly consistent with e.g.\ \cite{WahlEtAl2013}. We notice the somewhat higher trend since about 1998 in the reconstruction time series as compared to the tide gauges, which is likely due to merging the altimetry data into the spatial means of the reconstruction.

We
emphasize that creating the min-error $k$-OD triangulations took less than 10 milliseconds on average for $k\leq 3$, and for general min-error triangulations this required less than 100 milliseconds.  The implementation was done in Java and the experiments were performed on an Intel R Xeon R CPU E5-1620 processor. The machine is clocked at 3.6 GHz and has 32 GB RAM.
These running times suggest that our approach can also be applied on larger data sets taking far larger areas than the North Sea into account. 

\section{Conclusion} \label{sec:conclusion}
We have presented a novel and fast approach based on min-error $k$-OD  triangulations for reconstructing the dynamic sea surface. Our experiments give clear indications that the approach provides us with the possibility of reconstructing SSA maps with higher accuracy than those we obtain when using Delaunay triangulations. In particular, with our approach we were able to create reconstructions for the North Sea that span more than 20 years and still outperform the Delaunay triangulation. With this, we also offer evidence that integrating the altimeter data into the process of learning the reconstruction is worthwhile, rather than using it merely for evaluating the reconstructions.  

We suggest to use the min-error $2$-OD  triangulation, since it yields better reconstructions than the other min-error $k$-OD triangulations as well as the general min-error triangulation. Further, using this triangulation constitutes a good compromise between approximating the reference data and well-shaped triangles.

When it comes to choosing an appropriate training epoch for reconstructing the SSA map of a reconstruction epoch we suggest to use climatological triangulations, i.e., the training epoch stems from the same time in the year as the reconstruction epoch. This is a simple way to select a training epoch. However, other epochs might fit better to the reconstruction epoch. Future research could define a similarity score between the tide gauges of epoch $j$ and all available epochs $i$, and  choose the epoch with the greatest similarity score as training epoch.

We emphasize that the presented results of the experiments are only valid for the area of the North Sea. The next step is to test the method for other regions and to expand the method to areas of global extent.

\newpage

\section*{Data Availability}
The raw data can be downloaded at \cite{esa} (altimetry data) and at \cite{psmsl} (gauge data). For the processed data and the result data contact A. Nitzke (nitzke@igg.uni-bonn.de).

\section*{Computer Code Availability}
The source code of the presented algorithms called \emph{SeaSurfaceReconstruction} is available at \url{https://github.com/anitzke/seasurfacereconstruction}. Developer of the code is Alina F{\"o}rster. SeaSurfaceReconstruction is first available in 2020, the programming language is Java and the program size is about 285 kb. Java SE Development Kit 11 is required.
The contact address and e-mail of Alina F{\"o}rster are as follows:
\newline

\hspace*{0.5cm} Institute of Geodesy and Geoinformation\\
\hspace*{1cm} Working Group Geoinformation\\
\hspace*{1cm} Meckenheimer Allee 172\\
\hspace*{1cm} D - 53115 Bonn\\

\hspace*{0.5cm} nitzke@igg.uni-bonn.de\\

\bibliographystyle{styles/cas-model2-names}

\bibliography{triangulation}

\end{document}